\documentclass[aps,prl,twocolumn,amsmath,amssymb,floatfix,superscriptaddress,citeautoscript]{revtex4-2}

\usepackage{graphicx}
\usepackage{natbib}

\usepackage{color}
\usepackage{hyperref}
\usepackage{comment}

\begin{document}

	\title{Superconducting pairing symmetry in MoTe$_{2}$}
	
	\author{M. M. Piva}
	\email{Mario.Piva@cpfs.mpg.de}
	\affiliation{Max Planck Institute for Chemical Physics of Solids, N\"{o}thnitzer Stra{\ss}e 40, D-01187 Dresden, Germany}
	
	\author{L.~O.~Kutelak}
	\affiliation{Brazilian Synchrotron Light Laboratory (LNLS), Brazilian Center for Research in Energy and Materials (CNPEM), Campinas, Sao Paulo, Brazil}
	\affiliation{Max Planck Institute for Chemical Physics of Solids, N\"{o}thnitzer Stra{\ss}e  40, D-01187 Dresden, Germany}
	
	\author{R. Borth}
	\affiliation{Max Planck Institute for Chemical Physics of Solids, N\"{o}thnitzer Stra{\ss}e  40, D-01187 Dresden, Germany}
	
	\author{Y. Liu}
	\altaffiliation[Present address: ]
	{Los Alamos National Laboratory, Los Alamos, New Mexico 87545, USA.}
	\affiliation{Condensed Matter Physics and Materials Science Department, Brookhaven National Laboratory, Upton, New York 11973, USA}

	\author{C. Petrovic}
	\affiliation{Condensed Matter Physics and Materials Science Department, Brookhaven National Laboratory, Upton, New York 11973, USA}

	\author{R.~D.~dos Reis}
	\affiliation{Brazilian Synchrotron Light Laboratory (LNLS), Brazilian Center for Research in Energy and Materials (CNPEM), Campinas, Sao Paulo, Brazil}
	
	\author{M. Nicklas}
	\email{Michael.Nicklas@cpfs.mpg.de}
	\affiliation{Max Planck Institute for Chemical Physics of Solids, N\"{o}thnitzer Stra{\ss}e  40, D-01187 Dresden, Germany}

	\date{\today}
	
	\begin{abstract}
		
		Topological superconductors have long been sought for their potential use in quantum computing. The type-II Weyl semimetal MoTe$_{2}$ is an obvious candidate, exhibiting a superconducting state below 500~mK at ambient pressure, but the question remains whether the pairing is conventional $s^{++}$ or topological $s^{+-}$. The application of external pressure favors the superconducting state in MoTe$_{2}$ and suppresses the structural transition from $1T'$ to $T_{d}$. The competition between the two structures leads to a mixed phase that strongly enhances the disorder present in the system, remarkably without affecting the superconducting transition temperature, in contrast to the expectation of $s^{+-}$ pairing superconductivity. Our thorough analysis of the electrical and Hall resistivities as a function of pressure yields the most accurate temperature -- pressure phase diagram available to date for MoTe$_{2}$ and a detailed view of the relationship between disorder and superconductivity, supporting a conventional $s^{++}$ pairing symmetry.
		
	\end{abstract}

	\maketitle
	
	\newpage
	
	
	Much effort is devoted to the search for new materials that will enable the development of more efficient computing and storage systems  \cite{he2019topological}. In this regard, compounds or devices hosting Majorana fermions are being extensively sought. These particles are their own antiparticles, obey non-Abelian statistics, and have the potential to be used in fault-tolerant quantum computing \cite{he2019topological,sharma2022comprehensive,nayak2008nonabelian,lian2018topological}.
	
	Topological superconductors are promising materials to host Majorana fermions or Majorana zero modes. They exhibit gap-less excitations in their non-trivial surface states and an electron-pair condensate, which is a charge neutral ground state where electrons and holes are indistinguishable, allowing them to be their own antiparticles \cite{sharma2022comprehensive,wilczek2009majorana}. Topological superconductivity was predicted to be realized in fully gapped superconductors with odd parity symmetry, such as $p$-wave superconductors \cite{qi2011topological}, where Majorana fermions could exist in the vortex cores of these materials \cite{sharma2022comprehensive}. Recently it has been found that topological superconductivity is also possible in $s$-wave superconductors \cite{sato2010non}. For example, in time-reversal invariant Weyl semimetals with s$^{+-}$ superconductivity, a topological superconducting state might be present when the sign change of the superconducting gap occurs between Fermi surfaces of opposite Chern numbers \cite{hosur2014time}.
	
	The transition metal dichalcogenides are excellent candidates to host topological superconductivity. These materials have the chemical formula $TX_{2}$ ($T$ = transition metal, $X$ = chalcogenide) and many are predicted to be Dirac or Weyl semimetals \cite{sharma2022comprehensive}. Of particular interest is MoTe$_2$ \cite{Paul22}, which has three different structural phases \cite{guo2021recent}. A hexagonal $P6_{3}/mmc$ structure ($2H$ phase) is observed in semiconducting samples, but this phase can be easily tuned to the monoclinic $P2_{1}/m$ structure ($1T'$ phase), due to the low energy barrier between these structures. Finally, MoTe$_{2}$ can also exhibit a non-centrosymmetric orthorhombic $Pmn2_{1}$ structure ($T_{d}$ phase), which is stable at low temperatures or in thin films \cite{cui2019transport,he2018dimensionality,kowalczyk2023gate}. At ambient pressure, a first order structural transition is observed at $T_s\approx250$~K from the high temperature $1T'$ phase to the $T_d$ phase \cite{clarke1978low,guo2021recent,Paul22,2018_9113,qi2016superconductivity,hu2019angular,hu2020detection,liu2020quantum,2017_9124,Dissanayake19,yang2017elastic}. The variety of structural phases present in MoTe$_{2}$ and the small energy scales required to tune them suggest that structural disorder may play an important role in this material and that polymorphs might be kinetically stabilized, as is common in chalcogenides \cite{kanatzidis2017discovery}.
	
	In the $T_{d}$ phase, MoTe$_2$ is a type-II time-reversal invariant Weyl semimetal superconductor with a transition temperature ($T_{c}$) around 500~mK at ambient pressure \cite{wang2016mote,rhodes2017bulk}. The presence of both non-trivial topology and superconductivity has led to extensive studies investigating the possibility of topological superconductivity in MoTe$_{2}$ \cite{2016_11695}, however the superconducting pairing symmetry is still unknown. Muon spin rotation ($\mu$SR) \cite{guguchia2017signatures}, point contact spectroscopy \cite{luo2020possible}, and electronic gating experiments \cite{jindal2023coupled} revealed two band superconductivity, suggesting unconventional $s^{+-}$ or conventional $s^{++}$ pairing mechanisms, as observed in most of the iron pnictides \cite{fernandes2022iron} and MgB$_{2}$ \cite{szabo2001evidence}. A strong argument in favor of the $s^{+-}$ pairing is the suppression of $T_{c}$ with increasing disorder, as observed in MoTe$_{2}$ samples with different residual resistance ratios (RRR) \cite{rhodes2017bulk}. However, this cannot explain the enhanced $T_{c}$ observed in sulfur or Ta-doped samples \cite{li2018nontrivial,chen2016superconductivity,zhang2023enhanced}. Also $\mu$SR experiments cannot discriminate between $s^{+-}$ and $s^{++}$. Furthermore, other transport measurements have not detected a superconducting state down to 25~mK in high quality single crystals \cite{2018_9113}, while others claim that superconductivity resides in the $1T'$ phase \cite{lee2018origin}. Therefore, the realization of a superconducting non-trivial $s^{+-}$ phase in $T_d$ MoTe$_2$ is strongly debated.
	
	The application of external pressure suppresses the structural transition and strongly enhances $T_{c}$, as previously observed in many reports \cite{qi2016superconductivity,guguchia2017signatures,2018_9113,hu2019angular,hu2020detection,liu2020quantum}. However, a gray area remains in the temperature -- pressure ($T-p$) phase diagram of MoTe$_{2}$, the so-called mixed phase region, where both $T_{d}$ and $1T'$ structural phases are present. The existence of these different structural domains increases the disorder in the material, leading to higher residual resistivities and suppressed quantum oscillation amplitudes \cite{liu2020quantum,hu2020detection}. This enhancement of disorder may allow to distinguish between $s^{+-}$ and $s^{++}$ superconductivity in MoTe$_{2}$, as was previously demonstrated for iron pnictides \cite{ghigo2018disorder}.

	In this work we report a detailed study of the $T-p$ phase diagram of MoTe$_2$ using longitudinal and Hall resistivity measurements. Our data, taken at very small pressure steps, provide the most accurate $T-p$ phase diagram available to date for MoTe$_{2}$ and a detailed view of the relationship between structural phase transitions and superconductivity. A strong enhancement of the structural disorder in the mixed phase region was observed, remarkably without affecting the superconducting state, in contrast to the expectation for a $s^{+-}$ pairing state. Our findings shed new light on the pairing symmetry of the low pressure superconducting state of MoTe$_{2}$ and favor a conventional $s^{++}$ pairing mechanism.

	
	The electrical transport properties of single crystalline MoTe$_2$ grown from Te flux \cite{wang2016mote,yang2017elastic} were studied under hydrostatic pressure using a piston-cylinder type pressure cell with silicon oil as the pressure transmitting medium and lead as the pressure gauge \cite{nicklas2015pressure}. The magnetic field was applied parallel to the $c$-axis and the current was perpendicular to the field. Temperatures down to 50~mK and magnetic fields up to 9~T were achieved in different commercial cryostats.

	
	The $T-p$ phase diagram of MoTe$_2$ has been reported in several works before \cite{qi2016superconductivity,guguchia2017signatures,2017_9124,lee2018origin,2018_9113,2019_9095,hu2019angular,Dissanayake19,hu2020detection,liu2020quantum}, and for bulk samples a consensus has been reached that (i) at ambient pressure, a first-order type structural transition from the $1T'$ to the $T_{d}$ phase takes place around 250~K, which is suppressed to zero temperature in the vicinity of 1~GPa, and that (ii) at low temperatures, a superconducting phase is present in the entire temperature-pressure phase diagram up to 3 GPa and beyond. Only the early work of Qi {\it et al.}\ \cite{qi2016superconductivity} reported the presence of a structural transition up to about 3~GPa, which was not confirmed by later studies. This work has also led to the misconception that there is an almost instantaneous increase in $T_c$ within the $T_{d}$ phase below 1 GPa. All more recent studies show a smooth increase in $T_c$ \cite{guguchia2017signatures,2017_9124,lee2018origin,hu2019angular,hu2020detection}.

\begin{figure}[!t]
		\includegraphics[width=0.75\linewidth]{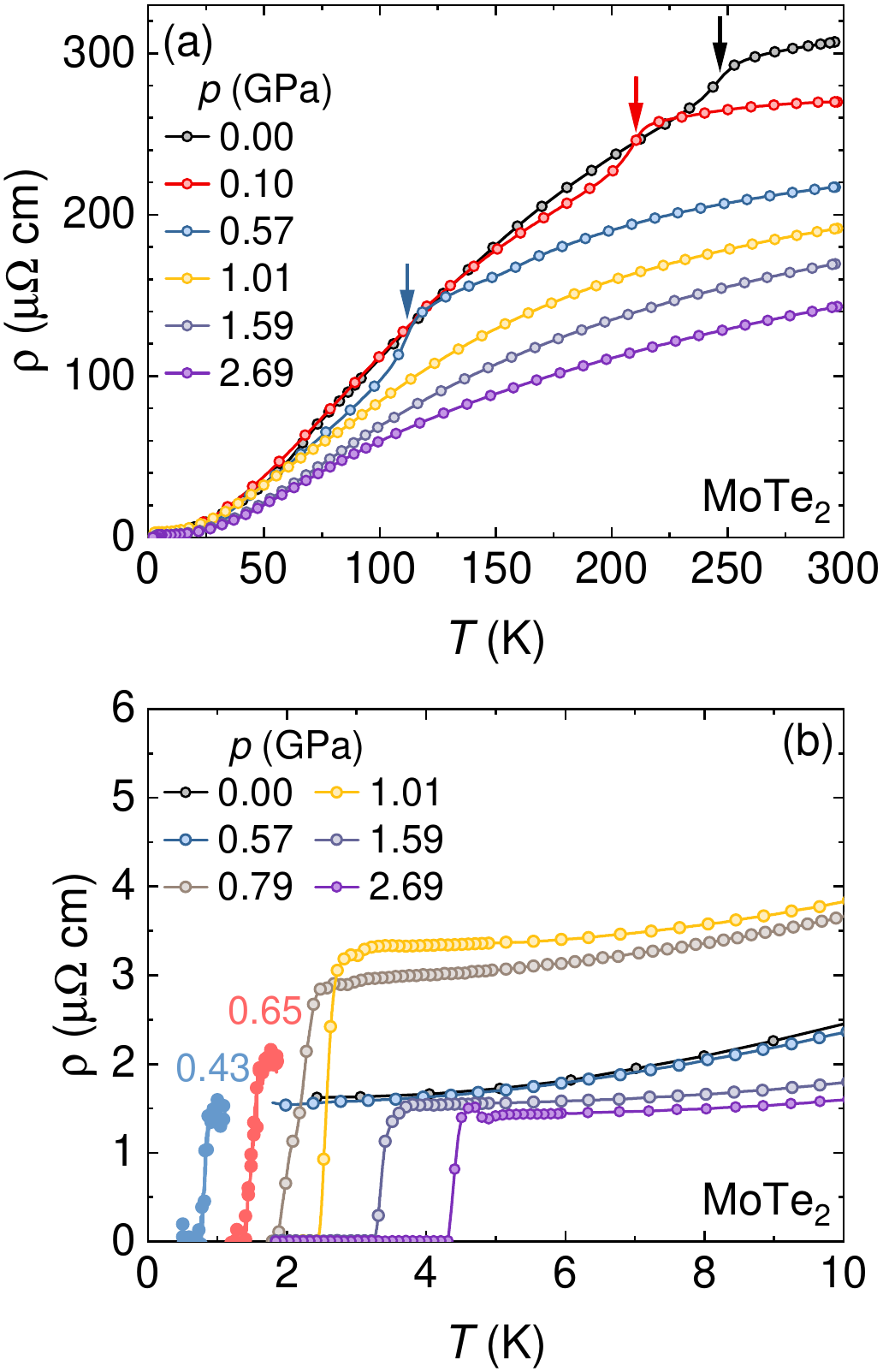}
		\caption{(a) Electrical resistivity ($\rho$) as a function of temperature for selected pressures. The arrows indicate the structural transition temperature. (b) Enlarged view of the low temperature region.}
		\label{Fig-res}
	\end{figure}

	The behavior summarized above is also observed in our study. We find an anomaly in the temperature dependence of the electrical resistivity $\rho(T)$ that characterizes the first-order structural transition from $1T'$ to $T_{d}$ at 246(2)~K at ambient pressure [see the black arrow in Fig.~\ref{Fig-res}(a)]. Increasing pressure suppresses the structural transition temperature $T_{s}$ to 53(5)~K at 0.87~GPa. Above this pressure, our $\rho(T)$ data show no evidence for the structural phase transition. Already at 0.93~GPa any feature is missing.

\begin{figure}[!t]
		\includegraphics[width=0.85\linewidth]{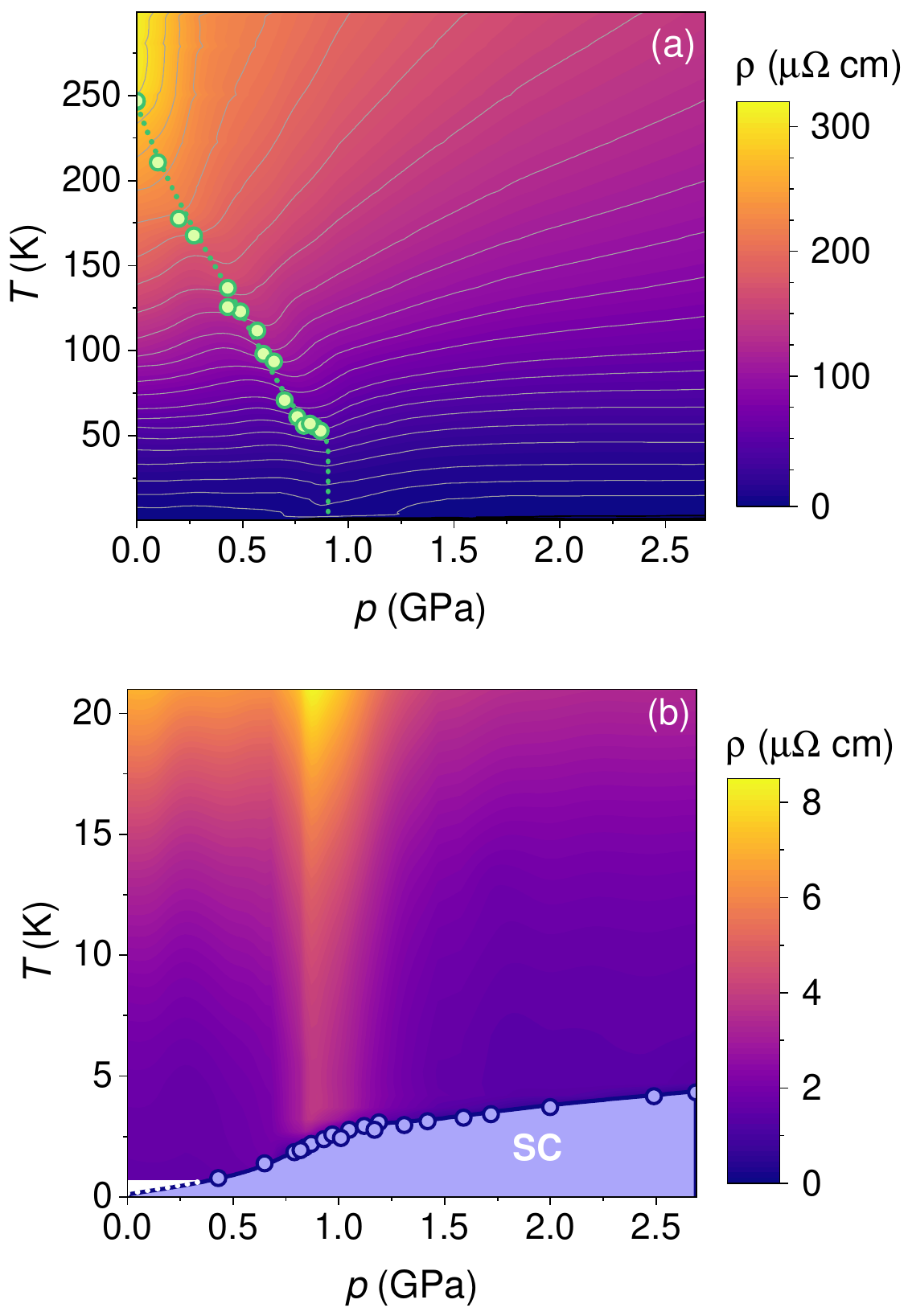}
		\caption{(a), (b) Color maps of the electrical resistivity ($\rho$) as a function of temperature for all studied pressures. Green and purple circles denote $T_{s}$ and $T_{c}$, respectively. The gray lines in (a) are equipotential lines of the resistivity and the dotted lines are guides for the eye. The data used to create the color maps are shown in the Supplemental Material \cite{SM} and are available at Ref.~\cite{EDMOND}.}
		\label{Fig-pd}
	\end{figure}

	To further investigate the first-order-like suppression of $T_{s}$, we performed temperature runs at closely spaced pressure points in the region around the critical pressure $p_c\approx0.9$~GPa. These data confirm the sudden suppression of $T_{s}$ and show a remarkable change in the curvature of $T_{s}(p)$. With increasing pressure, $T_{s}(p)$ changes from a concave to a convex curvature and shows a tendency to saturate before $T_{s}(p)$ is abruptly suppressed [see the phase diagram in Fig.~\ref{Fig-pd}(a)]. This behavior can be taken as a further indication of a first-order-like suppression of $T_{s}$. At low temperatures, we observe a strong variation of the normal state resistivity with pressure and zero resistance, indicating the development of a superconducting state [see Fig.~\ref{Fig-res}(b)]. Applying pressure increases the superconducting transition temperature $T_{c}$. We note that the superconducting transition remains narrow at all pressures. 
	
\begin{figure}[!t]
		\includegraphics[width=0.8\linewidth]{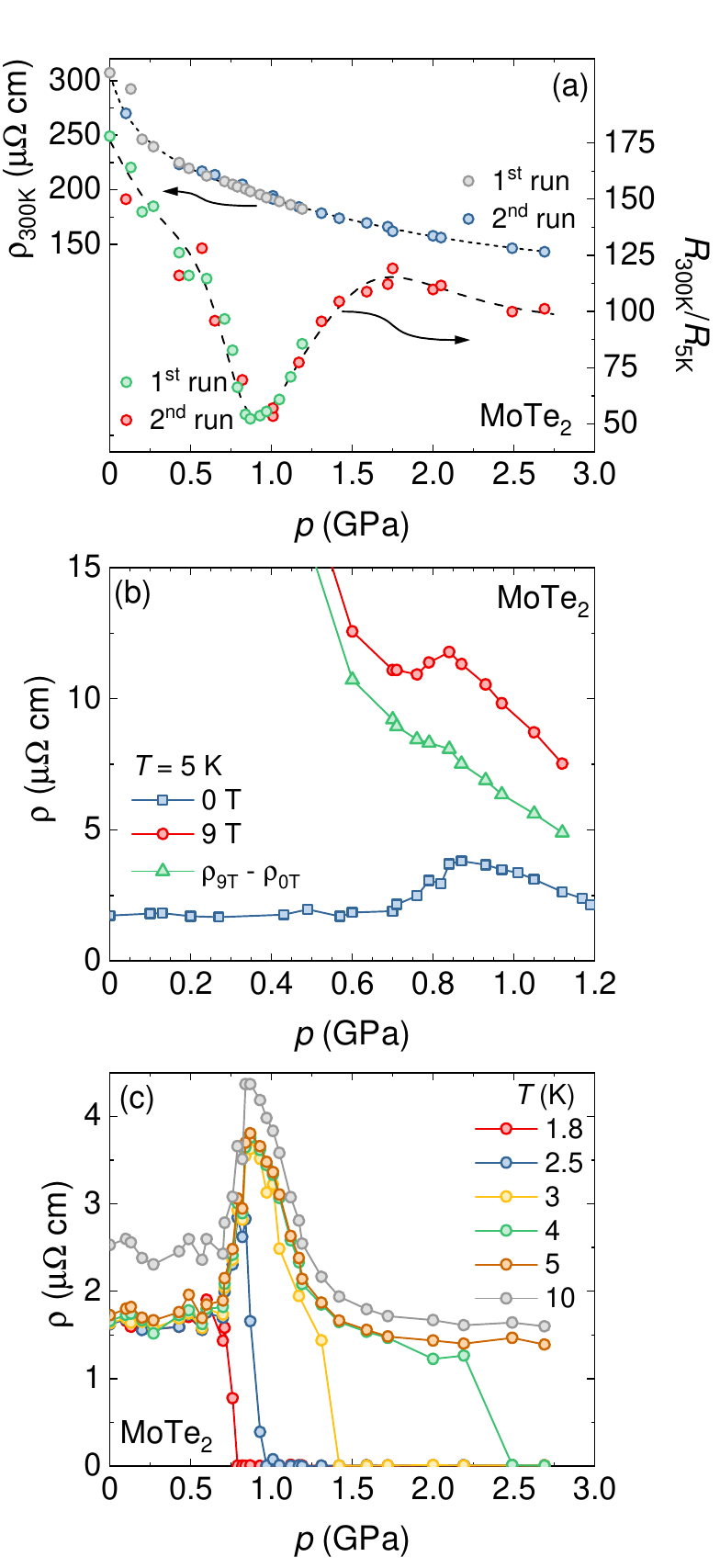} 
		\caption{(a) Left-axis, electrical resistivity at 300~K ($\rho_{300{\rm K}}$) and right-axis, residual resistivity ratio (RRR) as a function of pressure. The dashed and dotted black lines are guides to the eyes. (b) $\rho(p)$ at 5~K at 0 and 9~T and the difference between the two curves. (c) Electrical resistivity ($\rho$) in the low temperature range at selected temperatures as a function of pressure. }
		\label{Fig-res2}
	\end{figure}

	The residual resistivity ratio, ${\rm RRR}=\rho_{300{\rm K}}/\rho_{5{\rm K}}$, is the appropriate parameter to estimate the degree of disorder-related scattering present in a material. Figure~\ref{Fig-res2}(a) shows that the high-temperature resistivity $\rho_{300{\rm K}}$ decreases monotonically with pressure, but the RRR value is strongly suppressed in the mixed-phase region. The competition between the two structural phases leads to an enhancement of the structural disorder as well as the disorder-induced scattering in MoTe$_{2}$. The monotonic variation of $\rho_{300{\rm K}}(p)$ precludes any extrinsic origin. A clear indication of the structural origin of the enhanced disorder in the mixed-phase region is obtained by subtracting the electrical resistivity at 0 from the 9~T curve. As shown in Fig.~\ref{Fig-res2}(b), no resistivity anomaly is observed in the difference, ruling out a magnetic origin for the $\rho$ enhancement. Remarkably, the superconducting state is not affected by the increasing disorder in the mixed-phase region, as $T_{c}$ is monotonically enhanced with increasing pressure. Moreover, there is a sharp transition to the superconducting state that occurs independently of the pronounced changes in the resistivity value, as shown in Fig.~\ref{Fig-res2}(c).
	
	In the following we reconcile the implications of the structural transition, i.e.\ an inversion symmetry breaking transition from the centrosymmetric $1T'$ to the non-centrosymmetric $T_d$ phase, on the superconductivity, which has been proposed to be topological in the $T_d$ phase. $\rho_{\rm5K}(p)$, the isothermal resistivity at 5~K, is a good measure of the residual resistivity since there is little temperature variation in $\rho(T)$ at low temperatures $T\lesssim10$~K and the sample is in the normal state at this temperature throughout the pressure window studied here [see Fig.~\ref{Fig-res2}(c)]. It is pressure independent with a value of 1.7~$\mu\Omega$cm up to about 0.75~GPa, followed by a rapid rise to about 3.4~$\mu\Omega$cm at 1.01~GPa. Upon further pressure increase, $\rho_{\rm5K}(p)$ decreases again to a constant value of 1.4~$\mu \Omega$cm, which is slightly smaller than the ambient pressure value. The pressure independent behavior in the low- and high-pressure region away from the mixed phase suggests that pressure does not lead to any kind of structural defects and only drives the structural phase transition from the $T_{d}$ to the $1T'$ phase. The pronounced maximum in $\rho_{\rm5K}(p)$ is therefore directly related to additional scattering contributions in the mixed phase region and defines its range at low temperatures between about 0.75 and 1.4~GPa.

	\begin{figure}[!t]
		\includegraphics[width=1.00\linewidth]{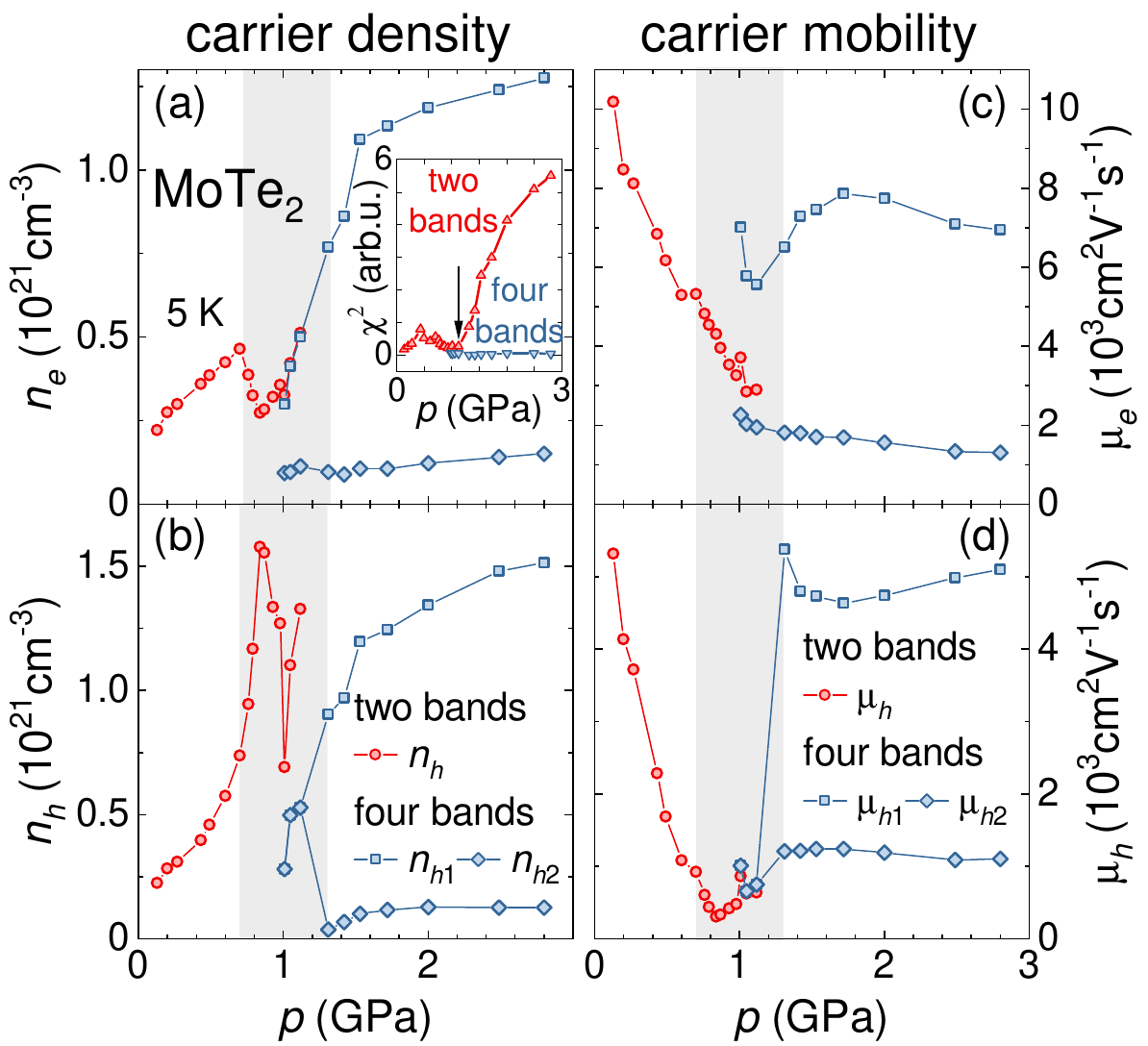}
		\caption{Carrier densities for (a) electrons and (b) holes and mobilities for (c) electrons and (d) holes for MoTe$_{2}$. The inset in (a) presents the reduced $\chi^{2}$ obtained with the fits considering two-band or four-band models. The highlighted area indicates the coexistence region of the two structural phases $T_{d}$ and $1T'$ occurs.}
		\label{Fig-carriers}
	\end{figure}
	
	To better visualize the features present in the electrical resistivity of MoTe$_{2}$, we show a color map in Fig.~\ref{Fig-pd}(a). At high temperatures the structural transition (green symbols) produces a distinct feature in the resistivity equipotential lines. Below about 50~K the feature does not move with pressure and remains at about 0.8 GPa. This indicates the first-order-like suppression of $T_{s}$, which is convincingly shown in the enlarged view of the low-temperature region in Fig.~\ref{Fig-pd}(b). The shading also indicates the region of increased structural disorder before the onset of superconductivity at lower temperatures. It is remarkable to see how the superconducting $T_c$ increases smoothly in this region. This behavior is not expected for two separate SC phases with different order parameter symmetry in the $1T'$ and $T_{d}$ structures and suggests the existence of a single superconducting phase in  MoTe$_{2}$.

	The analysis of the Hall conductivity gives more information about the carrier densities $n_{e(h)}$ and mobilities $\mu_{e(h)}$ (see the Supplemental Material \cite{SM} for further details). 
	Here $e(h)$ denotes electron(hole)-like charge carriers.  We obtain the longitudinal $\sigma_{xx}=\rho_{xx}/(\rho_{xx}^2+\rho_{xy}^2)$ and the transverse $\sigma_{xy}=-\rho_{xy}/(\rho_{xx}^2+\rho_{xy}^2)$ conductivities from our electrical and Hall resistivities. In MoTe$_2$, the application of external pressure enhances the transport contribution of four pockets (two holes and two electrons), as previously reported \cite{lee2018origin}. Therefore, the two-band model is more suitable for pressures lower than 1.12~GPa, while a four-band model has been applied for higher pressures, as clearly evidenced by the reduced $\chi^{2}$ obtained with the fits [see the inset of Fig.~\ref{Fig-carriers}(a)]. In the mixed-phase region, the results of the two models partly overlap. Figure \ref{Fig-carriers} summarizes the pressure evolution of the carrier densities and mobilities obtained from the transport data at 5 K for holes and electrons. At ambient pressure, the carrier density values are similar to previous reports \cite{hu2019angular}. The highlighted gray area represents the mixed-phase region. One can see a smooth increase in the carrier densities of both electrons and holes with increasing pressure, which then starts to decrease in the mixed-phase region, reaching a minimum at 0.84~GPa [see Figs.~\ref{Fig-carriers}(a) and \ref{Fig-carriers}(b)]. Above this pressure the carrier densities start to increase again and saturate at high pressures for all 4 pockets. Both electron and hole mobilities are suppressed with increasing pressure and reach a minimum in the mixed-phase region [see Figs.~\ref{Fig-carriers}(c) and \ref{Fig-carriers}(d)]. After the appearance of the new hole and electron pockets, the mobility for one pocket is almost pressure independent, while for the others it is strongly increased and then saturates above 1.5~GPa. These results highlight the distinct differences in the electronic properties of the $1T'$ and $T_d$ phases, in contrast to the smooth evolution of the superconducting $T_c$ as a function of pressure over the entire pressure range studied.

	Our results show a smooth variation of $T_c(p)$, independent of the large increase in the residual resistivity, indicating enhanced disorder scattering, near the critical pressure where $T_s$ is suppressed. Furthermore, pronounced changes in the electronic properties were evidenced by the distinct pressure evolution of the charge carrier density and mobility. These features enabled us to draw conclusions about the possible superconducting order parameter(s) in MoTe$_{2}$.
	A topological $s^{+-}$ superconducting state has been proposed to be realized in the non-centrosymmetric $T_d$ phase and a conventional $s^{++}$ state in the centrosymmetric $1T'$ phase \cite{guguchia2017signatures,guguchia2020pressureinduced}. However, it is quite challenging to directly probe the relationship between the superconducting state and the crystalline structure as a function of pressure at temperatures below 5~K.
	While $s^{++}$ superconductivity is nearly insensitive to structural disorder, $s^{+-}$ is expected to be strongly suppressed by it \cite{hosur2014time,qi2011topological}. We observe an enhancement of $T_c$ with increasing pressure in the mixed-phase regime where the residual resistivity is maximal. This seems to rule out the possibility of topological $s^{+-}$ pairing in MoTe$_{2}$. We therefore propose a scenario in which only the $1T'$ phase exhibits (conventional) superconductivity at low temperatures and the $T_d$ phase remains normal at all pressures. 
 
 Previous studies proposing a topological $s^{+-}$ pairing in MoTe$_{2}$ based on the suppression of $T_c$ as a function of disorder did not consider the possible presence of the $1T’$ phase at low temperatures in the entire pressure range down to ambient pressure. Our proposal implies that a small amount of the $1T'$ phase is already present in the material at low pressures. Cho {\it et al.}\ actually report the presence of the $1T'$ phase at ambient pressure as a function of the residual resistivity ratio \cite{2017_9122}. According to this work, the ${\rm RRR}= 178$ of our sample would imply a very small fraction of the $1T'$ phase of a few percent present at ambient pressure. This is not inconsistent with other work claiming only the presence of the $T_d$ phase at low temperatures, since such a small volume fraction of the $1T'$ phase may not be resolved in X-ray diffraction. Evidence for the presence of the $1T'$ phase at low pressures has been reported in Ref.~\cite{2018_9113,2017_9122} and no zero-resistance state has been observed in samples lacking the $1T'$ structure. We note that even a small volume fraction of the $1T'$ phase in the sample could cause a sharp superconducting jump in resistivity and produce a full-volume diamagnetic shielding in AC susceptibility experiments, as previously reported for polycrystalline samples \cite{guguchia2017signatures}. The smooth evolution of $T_c(p)$ is also difficult to reconcile with an expected phase transition from a topological $s^{+-}$ to a conventional $s^{++}$ superconducting state. Our proposal of a (conventional) superconducting state in the $1T'$ phase, present over the entire pressure range at low temperatures, explains the observed behavior in a natural way.

	In conclusion, we have performed a detailed investigation of the temperature -- pressure phase diagram of MoTe$_{2}$ by electrical transport experiments. Our data reveal a first-order-like suppression of the structural phase transition and a strong increase of the disorder scattering contribution to the electrical resistivity in the mixed-phase regime around the first order phase transition. Surprisingly, we find a smooth increase of the superconducting transition temperature over the whole pressure range. This leads us to conclude that topological $s^{+-}$ superconductivity is most likely not realized in MoTe$_{2}$. Based on our data, we argue that only the $1T'$ phase exhibits superconductivity at low temperatures.
	
	Data that underpin the ﬁndings of this paper are available at Edmond – the open research data repository of the Max Planck Society at Ref.~\cite{EDMOND}.
	
	
	\begin{acknowledgments}
		We acknowledge fruitful discussions with Michael Baenitz, Takuto Fuji and Hiroshi Yasuoka. This project has received funding from the European Union’s Horizon 2020 research and innovation programme under the Marie Sk\l{}odowska-Curie grant agreement No 101019024. This work was also supported by the S\~ao Paulo Research Foundation (FAPESP) grants 2018/0823-0, 2021/02314-9 and 2022/05447-2. Work at Brookhaven National Laboratory was supported by U.S. DOE, Office of Science, Office of Basic Energy Sciences under Contract No. DE-SC0012704 (materials synthesis). R.D.d.R. and L.O.K. acknowledge financial support from the Max Planck Society under the auspices of the Max Planck Partner Group R. D. dos Reis of the Max Planck Institute for Chemical Physics of Solids, Dresden, Germany.
	\end{acknowledgments}
	
	\bibliography{MoTe2}

	\begin{figure*}[!t]
	
	\centering
	\includegraphics[width=1\textwidth]{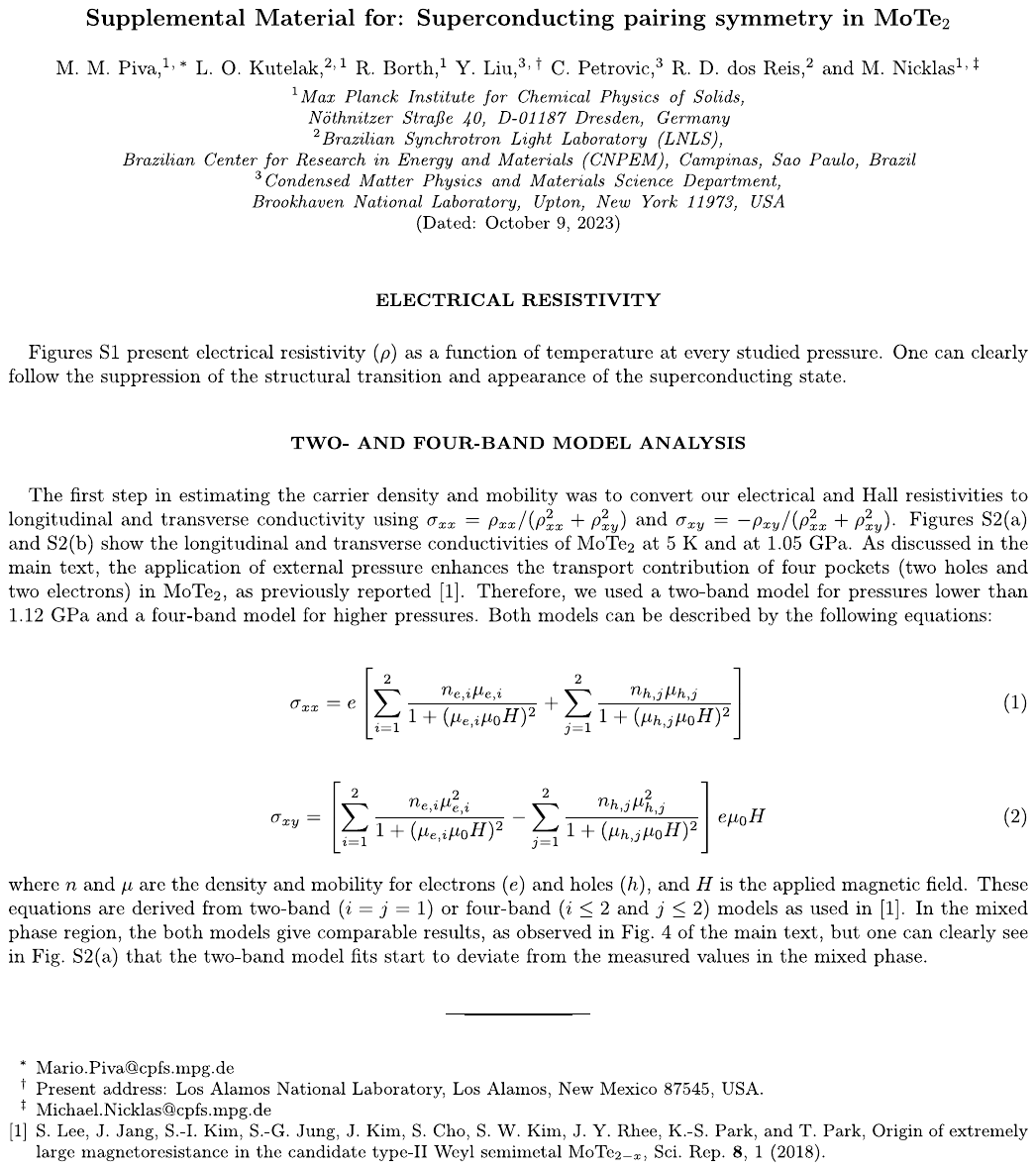}
\end{figure*}

	\begin{figure*}[!t]
	
	\centering
	\includegraphics[width=1\textwidth]{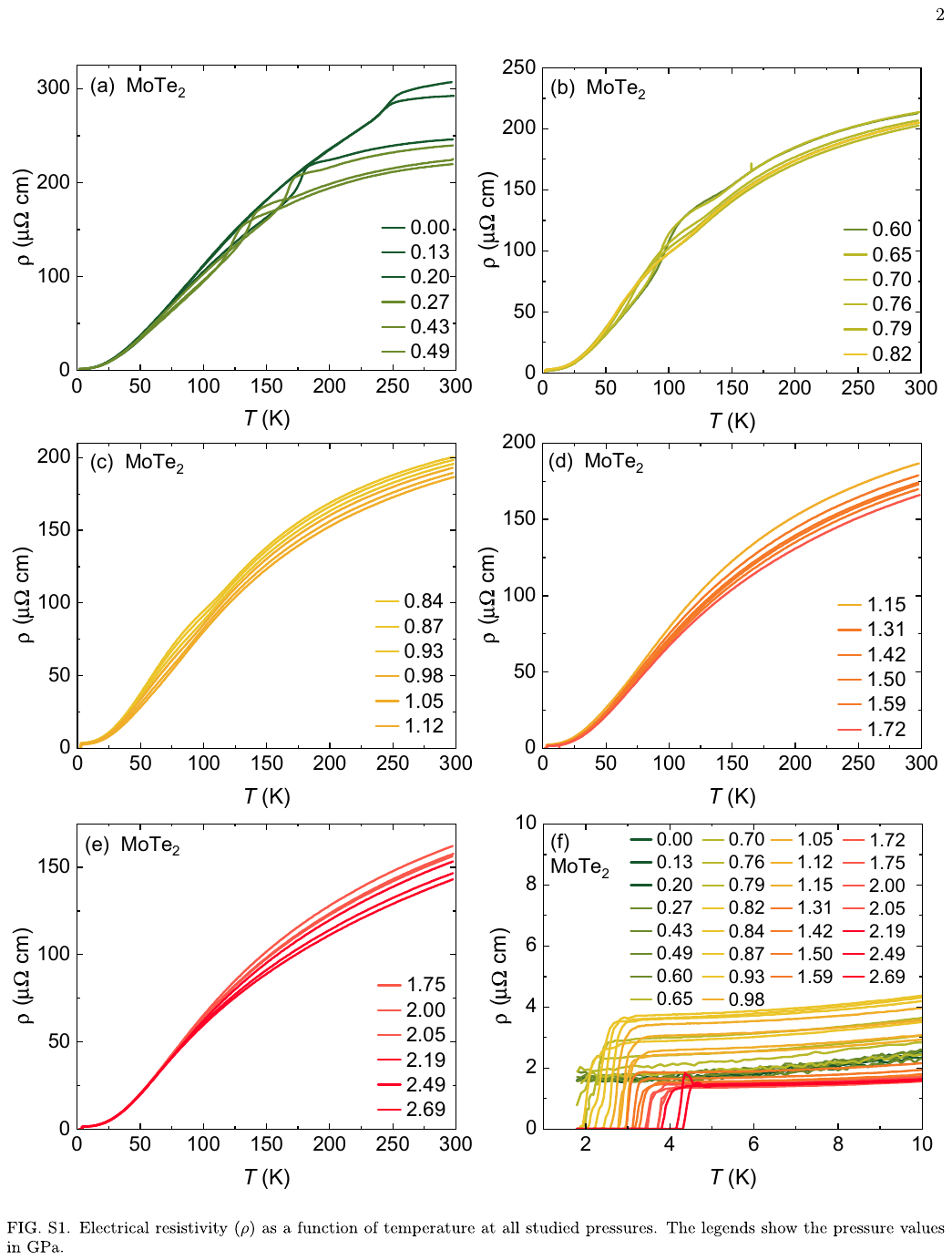}
\end{figure*}

	\begin{figure*}[!t]
	
	\centering
	\includegraphics[width=1\textwidth]{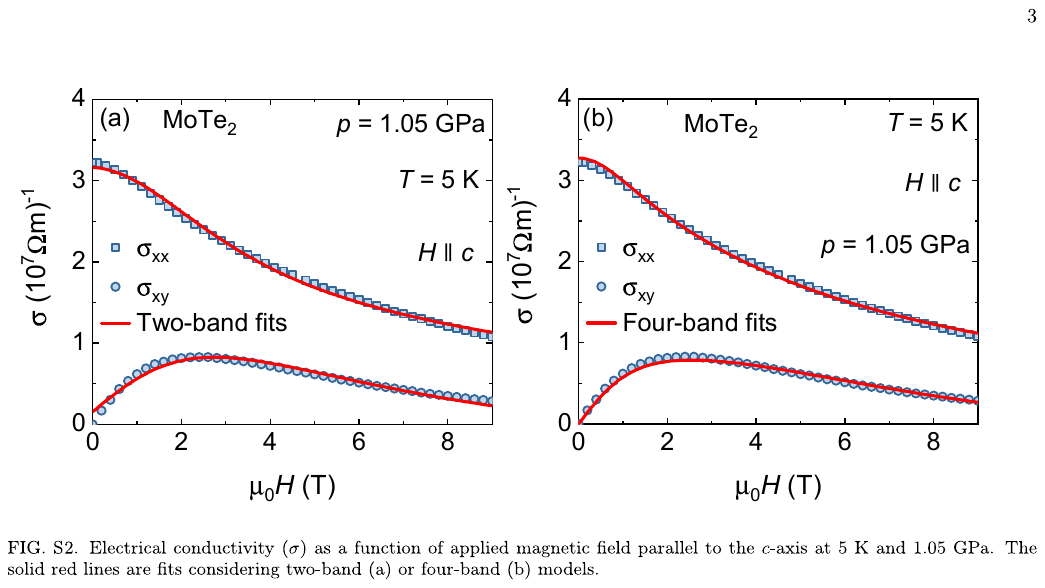}
\end{figure*}

\end{document}